\documentclass[useAMS,usenatbib]{mn2e}
\usepackage{epsfig,graphicx,latexsym,amsmath,amssymb}
\usepackage{natbib}
\usepackage{hyperref}
\usepackage{mathrsfs}
\usepackage{lastpage}
\usepackage{bm}
\usepackage[labelformat=empty]{subfig}

\def\msun{{\,{M}_\odot}}
\def\yrs{{\,{\rm yr}}}
\def\simlt{\lower.5ex\hbox{$\; \buildrel < \over \sim \;$}}
\def\simgt{\lower.5ex\hbox{$\; \buildrel > \over \sim \;$}}
\newcommand {\PN}{{\rm PN}}

\bibpunct[,]{(}{)}{;}{a}{}{,}

\title[The Schwarzschild barrier]
      {Blocking low-eccentricity EMRIs:\\
       A statistical direct-summation $N-$body study of the Schwarzschild barrier}

\author[Patrick Brem, Pau Amaro-Seoane, Carlos F. Sopuerta] % Running authors, 
{Patrick Brem$^{1}$
                        \thanks{E-mail: Patrick.Brem@aei.mpg.de}, 
Pau Amaro-Seoane$^{1}$ \& Carlos F. Sopuerta$^{2}$
   \\
$^{1}$Max Planck Institut f\"ur Gravitationsphysik
(Albert-Einstein-Institut), D-14476 Potsdam, Germany
\\
$^{2}$Institut de Ci\`encies de l'Espai (CSIC-IEEC), 
Facultat de Ci\`encies, Campus UAB, Torre C5 parells, 
08193 Bellaterra, Spain
}

\begin{document}

\date{draft \today}

\pagerange{\pageref{firstpage}--\pageref{lastpage}} \pubyear{2012}

\maketitle

\label{firstpage}

\begin{abstract}
The capture of a compact object in a galactic nucleus by a massive black hole
(MBH), an extreme-mass ratio inspiral (EMRI), is the best way to map
space and time around it.  Recent work on stellar dynamics has demonstrated
that there seems to be a complot in phase space acting on {\em
low-eccentricity} captures, since their rates decrease significantly by the
presence of a blockade in the rate at which orbital angular momenta change
takes place. This so-called ``Schwarzschild barrier'' is a result of the impact
of relativistic precession on to the stellar potential torques, and thus it
affects the enhancement on lower-eccentricity EMRIs that one would expect from
resonant relaxation. We confirm and quantify the existence of this barrier
using a statistical sample of 2,500 direct-summation $N-$body simulations using
both a post-Newtonian and also for the first time in a direct-summation code a
geodesic approximation for the relativistic orbits.  The existence of the
barrier prevents low-eccentricity EMRIs from approaching the central MBH, but
high-eccentricity EMRIs, which have been wrongly classified as ``direct
plunges'' until recently, ignore the presence of the barrier, because they are
driven by two-body relaxation.
Hence, since the rates are significantly affected in the case of low-eccentricity
EMRIs, we predict that a LISA-like observatory such as eLISA will predominantly
detect high-eccentricity EMRIs.
\end{abstract}


\section{Introduction}
\label{sec.intro}

Massive black holes (MBHs), with masses ranging from some $10^4\,M_{\odot}$ to
a few $10^9\,M_{\odot}$ are very likely present in the centre of most galaxies.
Measurements of the kinematics of gas and stars in the central regions of
nearby galaxies \citep[see e.g.][]{deZeeuw00,Barth04,Kormendy04,Richstone04}
have provided us with compelling evidence.  Our own Milky Way (MW) is the
galaxy for which we have the strongest observational proof a central MBH. Data
based on 16 years of observations set the mass of the central SMBH to $\sim 4
\times 10^{6} \, M_{\odot}$
\citep{EisenhauerEtAl05,GhezEtAl05,GhezEtAl08,GillessenEtAl09,GenzelEtAl10}.

To interact with the central MBH, stars have to find themselves on
``loss-cone'' orbits, which are orbits elongated enough to have a very close-in
pericentre \citep{FR76,LS77,AS01}.  While main-sequence stars are tidally
disrupted when approaching the central massive black hole (MBH), compact
objects (stellar black holes, neutron stars, and white dwarfs) slowly spiral
into the MBH via the gradual loss in the form of gravitational radiation and
are eventually swallowed after some $\sim 10^{3-5}$ orbits in the eLISA
band\citep{Amaro-SeoaneEtAl07,Amaro-SeoaneEtAl2012,Amaro-SeoaneEtAl2012b,Amaro-SeoaneLRR2012}.
This is the best way to probe general relativity \citep{Sopuerta2010} and,
thus, the factory of space and time around a massive black hole.  We will also
get additional information about the binary itself: in particular the masses
of the system and the spins of the MBH can be measured to 
a level of precision without any precedent 
\citep[see][and references therein]{Amaro-SeoaneEtAl07}. Besides, the
detection will provide us with information about the distribution of dark
objects in galactic nuclei and globular clusters. 

Producing EMRIs is more difficult than producing tidal disruptions of stars
(e.g., \citealt{Rees88,MT99,SU99,WM04}) because while disruptions require a
single passage within a critical radius, an EMRI is a progressive phenomenon
that is only successful if the small compact object suffers a large number of
close encounters with the central MBH.

The requirement for an EMRI to be successful is not just to have a small
periapsis, but that the gravitational radiation timescale is sufficiently
shorter than 2-body relaxation, which could have a significant impact on the
periapsis. Although the basic ideas are explained in detail in the recent
review of \cite{Amaro-SeoaneLRR2012}, we deem it important to briefly summarize
in the next paragraphs the fundamentals of the ideas of two different
categories of EMRIs, since this will be crucial to understand the main results
of this paper.

In a galactic nucleus {without dissipation processes} a star suffers
``gravitational tugs'' in the regime where the evolution is dominated by close
encounters with other stars. These tugs, since they are driven by two-body
relaxation, are {\em random} and originated by interacting with other stars
that happen to have a very close position. The scattering rate is very similar in
orbital energy $E$ and angular momentum $L$.  If the star gets close to a very low $L$,
which is statistically probable, then the picture changes: The rate at which
the star changes $L$ will be much shorter than that at which it changes $E$. If
we introduce a dissipation term in the picture, e.g. gravitational waves, the
star follows the same initial evolution and, at some point, our test star, the
compact object, reaches the region in which it is on a very radial orbit.  In
this case, at every periapsis passage, the test star loses a significant amount
of energy and, hence, the semi-major axis shrinks. If the process is
efficient enough it becomes an EMRI.

Stars in very radial orbits would then scatter faster in $L$ than in $E$, so
that they would end up as direct plunges, i.e. being swallowed by the MBH after
an insignificant amount of gravitational wave bursts. Although their event rate
is much larger, they do not allow us to get a full picture of the scenario as
complete as compared with the slow inspiral of an EMRI \citep[but see the
recent work of][on bursting sources]{BerryGair2012}.  The threshold lies around
$\sim 10^{-2}$ pc \citep{HopmanAlexander05} and this is what has led up to now
to the thought that the EMRI event rate should be dominated by the physical
phenomena happening in the innermost volume around the MBH, of radius $\sim
10^{-2}$ pc.

However, while this is strictly true for Schwarzschild MBHs, {\em the situation for
spinning MBHs drastically changes the narrative}. Recently,
\cite{Amaro-SeoaneSopuertaFreitag2012} proved that for Kerr MBHs plunges do not
plunge, but spend a very large number of cycles in the eLISA band. I.e. they
are simply high-eccentricity EMRIs. The authors prove that the event rate of
both high-eccentric and also low-eccentric EMRIs is enhanced by the spin as
compared to the Schwarzschild case by an amount that depends on the specific
eccentricity and inclination of the orbit.

The fact that compact objects on a ``plunge'' orbit have been envisaged as
uninteresting has led to an effort to understand the phenomena that could lead
to the creation of EMRIs in a volume of radius $\sim 10^{-2}$ pc. In this
volume, resonant relaxation is very likely the most important process to lead
compact objects to EMRI orbits \citep{HopmanAlexander06}. Nonetheless, while in
the absence of relativistic effects resonant relaxation is expected to change
the angular momentum of the stellar BHs very efficiently, \cite[][ hereafter MAMW11]{MerrittEtAl11}
showed recently with a few direct $N-$body experiments  that introducing
relativistic precession effects are both a blessing and a curse for the
inspiral event rate: RR is quenched at high eccentricities, resulting in more
inspirals than plunges. However, this quenching also means that, in total,
fewer BHs will reach pericentres that are small enough to lead to an inspiral
observable by gravitational wave detectors such as eLISA 
\citep{Amaro-SeoaneEtAl2012,Amaro-SeoaneEtAl2012b}. In this paper we present a
statistical study of the Schwarzschild barrier (SB) with a set of 2,500
direct-summation $N-$ body simulations including relativistic corrections to
study and quantify this effect.  We implement the relativistic effects using a
post-Newtonian formalism as in \cite{KupiEtAl06} but also, and {\em for the first
time ever in a direct-summation integrator, a geodesic scheme}.

This paper is organized as follows: In Section \ref{sec.methods} we present the
physical setup and the numerical methods used. In Section \ref{sec.results} we
present general results and discuss the implications for the Schwarzschild barrier in Section
\ref{sec.sb}.

\section{Physical setting and numerical methods}
\label{sec.methods}

Recently, MAMW11 estimated with a few direct-summation N-body
simulations expanded with a statistical Monte-Carlo study
that the traditional EMRI event rate is markedly decreased by
the presence of a blockade in the rate at which orbital angular
momenta change takes place. This so-called Schwarzschild
barrier is a result of the impact of relativistic precession
on to the stellar potential torques. Although the authors
find that some particles can penetrate the barrier, EMRIs
are significantly suppressed in this scenario.

In analogy to MAMW11, the setup we consider consists of a central MBH
of mass $M_\bullet$ surrounded by 50 stellar mass black holes (BHs) of mass $m_\star$. 
We fix the masses of the MBH and the stellar black holes at 

\begin{align}
  M_\bullet &= 10^6 \msun \,,\nonumber \\
  m_\star &= 50 \msun\,.
\end{align}

\noindent
Initially we distribute the stars in phase space following
a distribution of the form $N(a,e^2)\,da\,de^2 = N_0 da\,de^2$, with $a$ the semi-major 
axis of the BHs and $e$ their eccentricity. The semi-major axes 
range between $0.1~{\rm mpc} < a < 10~{\rm mpc}$.

This setup represents roughly a relaxed distribution of stellar mass objects around a MBH \citep{FAK06a}. 
We note that the event rates we obtain in this study are only applicable to this specific, idealized system 
where $N=50$ heavy, equal mass stellar BHs orbit a MBH. For more realistic estimates one would need to take 
into account other properties of galactic nuclei, such as the stellar background, a certain mass distribution 
and larger $N$.

\subsection{Timescales}
In order to have EMRI events, one needs BHs on orbits with pericentres of only a few gravitational radii
($r_g = GM_{\bullet}/c^{2}$).  This requires the existence of physical mechanisms for driving BHs from 
their initial orbits onto very eccentric ones.

A purely Newtonian system has two different ways of exchanging angular momentum $L$ (or eccentricity $e$) and 
orbital binding energy $E$ (or semi-major axes $a$).  The first one is by two-body scattering, or non-resonant 
relaxation (NR) \citep[see e.g.][]{Spitzer87,BinneyTremaine08,Amaro-SeoaneLRR2012}. 
Every time two objects come close they undergo scattering, 
changing the momenta of the scattering partners. This very basic mechanism exists in all gravitationally 
interacting systems from compact star clusters to galaxies. The associated time-scale for changing the angular 
momentum $L$ of a given particle by $\Delta L \sim L$ is

\begin{equation}
\tau_{\rm NR} = 4.6\,{\rm Myr}\,\tilde{a}^{1/2} \Bigl(\frac{M_{\bullet}}{10^6 \msun}\Bigr)^{3/2} 
\Bigl(\frac{m}{50 \msun}\Bigr)^{-2}\Bigl(\frac{N_{<1}}{5}\Bigr)^{-1},
\label{eq.nr}
\end{equation}
where $N_{<1}$ is the number of stars within a sphere of $1\;$mpc and $\tilde{a}$ will denote the 
semi-major axis in mpc throughout the rest of the paper. The derivation of this equation, 
although trivial, can be found in MAMW11.

In Newtonian systems with a central massive object which dominates the gravitational potential, 
however, this relaxation mechanism is usually dominated by resonant relaxation (RR) \citep{HopmanAlexander06}. 

The torque of a spherical distribution of stars of mass $M_\star(r < a)$ inside a sphere of radius 
$r$ on a BH with semi-major axis $a$ leads to a retrograde orbital in-plane precession on a time-scale

\begin{equation}
\tau_{\rm M} = \frac{2 \pi}{g(e) M_\star(r < a)}\Bigl(\frac{M_\bullet a^3}{G}\Bigr)^{1/2},
\label{eq.m}
\end{equation}
where the eccentricity dependent function $g(e)$ is given by

\begin{equation}
g(e) = \frac{1 + \sqrt{1-e^2}}{2\sqrt{1-e^2}}\,.
\end{equation}
>From Eq.~(\ref{eq.m}) we can derive the time-scale of changes in the angular momentum, 
which for our system is given by

\begin{equation}
\tau_{\rm RR} \approx \frac{5.9 \times 10^4 \,{\rm yr}}{\beta_s^2 g(e)}
\Bigl(\frac{M_\bullet}{10^6 \msun}\Bigr)^{1/2}\Bigl(\frac{m_\star}{50\msun}\Bigr)^{-1} \tilde{a}^{3/2}\,,
\label{eq.rr}
\end{equation}
where $\beta_s$ is a factor of order unity.
For the derivation of Eq.~(\ref{eq.rr}) we defer the reader to reference MAMW11.
This mechanism for relaxation, RR, is much more efficient than NR {\em at these distances} because the particles interact 
through coherent torques in resonant Keplerian orbits. It therefore could lead in principle to an enhancement in the EMRI
event rate.

Relativistic effects introduce two new time-scales. The conservative Schwarzschild precession, 
appearing in particular at the first and second post-Newtonian orders, causes a precession of the 
pericentre by an angle

\begin{equation}
\delta \Phi = \frac{3 \pi G M_\bullet}{c^2} \frac{1}{a (1-e^2)}
\end{equation}
per orbit. This leads to the following associated time-scale 

\begin{equation}
\tau_{\rm SS} = \frac{\pi}{\delta \Phi}  P(a) = \frac{2 \pi c^2}{3 (G M_\bullet)^{3/2}} a^{5/2} (1-e^2),
\label{eq.ss}
\end{equation}
where $P(a) = 2 \pi (a^3/GM_{\bullet})^{1/2}$ is the orbital period. In a more convenient notation this yields

\begin{equation}
\tau_{\rm SS} \approx (2 \times 10^4 \yrs) \, \tilde{a}^{5/2} (1-e^2) \Bigl(\frac{M_\bullet}{10^6 \msun}\Bigr)^{-3/2}.
\end{equation}
The second important time-scale is the inspiral time $\tau_{\rm GR}$ via gravitational radiation only 
which, for high eccentricities ($e\simeq 1$), 
is given by \citep{Peters64} 

\begin{align}
 \tau_{\rm GW} \approx & \frac{5 c^5}{256 G^3}\frac{a^4}{m_\star M_\bullet (m_\star + M_\bullet)}(1-e^2)^{7/2}\\
\nonumber \approx & (1.16 \times 10^{13} \yrs)\, \tilde{a}^4 (1-e^2)^{7/2} \\
&\times\Bigl(\frac{m_\star}{50 \msun}\Bigr)^{-1} \Bigl(\frac{M_\bullet}{10^6 \msun}\Bigr)^{-2}\,,
\label{eq.peters}
\end{align}
for $M_\bullet \gg m_\star$. This timescale is highly sensitive to the eccentricity and semi-major 
axis and for a typical BH in the system much longer than any other relevant time. However, for 
particles very close to the central MBH, gravitational radiation may drive them gradually into 
the capture radius leading to an ``inspiral event''.

\subsection{A direct-summation code with post-Newtonian and geodesic corrections}

In order to integrate the initial configuration over time we use the 
publicly available {\tt planet} code by Sverre Aarseth~\citep{Aarseth99,Aarseth03}, a direct summation $N-$body 
integrator.  We have modified this code in order to introduce relativistic 
corrections to the Newtonian acceleration \citep{Amaro-SeoaneBremCuadraArmitage2012}.  For the studies that we present here
we have considered the following types of dynamics:

\begin{itemize}
  \item purely Newtonian dynamics
  \item post-Newtonian (PN) corrections
  \item relativistic geodesic equations for motion of the particles around the MBH
\end{itemize}

In the purely Newtonian case, the integration is obviously done without modifications to the 
acceleration equations. In the second case we add the PN corrections in the following way:

\begin{equation}
{F}  = \overbrace{{F}_0}^{\rm Newt.}
+\overbrace{\underbrace{c^{-2}{F}_2}_{1\PN} +
\underbrace{c^{-4}{F}_4}_{2\PN}}^{\rm periapsis~shift} +
\overbrace{\underbrace{c^{-5}{F}_5}_{2.5\PN}}^{\rm energy~loss} +
\overbrace{\mathcal{O}(c^{-6})}^{\rm neglected}\,,
\label{eq.F_PN}
\end{equation}
where the individual $F_i$'s denote the different PN corrections to the total force on
a particle, which can be found in Appendix~\ref{pn-corrections}. 

Given the high mass ratios involved in EMRIs, their motion around a MBH can be also approximated 
by solving the geodesic equations of motion, neglecting in this way dissipative effects due
to  gravitational wave emission and higher-order corrections in the mass ratio.  
In our case, the geodesic equations describe the exact trajectory of a test mass 
particle around a Schwarzschild MBH.  {\em Unlike the PN approximation, the geodesic equations are 
valid even in the last few $r_g$ during a plunge or inspiral}, however only in the limit 
$m_\star/M_\bullet \rightarrow 0$. Some orbits are expected to migrate towards plunge or 
inspiral orbits at pericentre distances of $r_p < 15 \,r_g$, where the errors of the PN 
approximation can already be quite significant \citep{YunesBerti2008}. In order to test the 
existence of the Schwarzschild barrier at small distances, we have implemented these corrections 
in the {\tt planet} code.

Since the geodesic equations do not contain dissipative terms we compare the 
results of using them with the conservative PN implementations, i.e. setting the dissipative 
correction $F_5 = 0$ in Eq.~(\ref{eq.F_PN}).

At any given time all the active acceleration corrections are computed only between the 
MBH and a stellar BH object.  The semi-major axis and eccentricity evolution is tracked by 
monitoring the distances at periapsis $r_{\rm min}$ and apoapsis $r_{\rm max}$ using the 
standard relations

\begin{align}
  a &= \frac{r_{\rm max} + r_{\rm min}}{2} \,, \\
  e &= \frac{r_{\rm max} - r_{\rm min}}{r_{\rm max} + r_{\rm min}}\,,
\end{align}
which are valid for any acceleration correction. This calculation does not require a PN expansion 
of the Keplerian expressions for $a$ and $e$ and is thus consistent with the purely Newtonian, PN and 
geodesic equations of motion.

We record a merger event whenever a particular has an instantaneous separation to the MBH of $r < 6 \, r_g$
(i.e. the Schwarzschild last stable orbit for circular orbits). We note that MAMW11 use $r < 8\,r_g$ 
and thus might classify a few events that we call inspirals as plunges.
We run $N_{P} = 500$ simulations of the models described above, with different random 
seeds for the distribution of stars, for a few Myr or until a merger event happens.

\section{Results from the Simulations}\label{sec.results}

For all the different sets $P$ of simulations we will compute the average time $\tau_{X,P}$ for 
the occurrence of a certain event $X$ in the simulation set $P$ in the following way:

\begin{equation}
\tau_{X,P} = \sum_{i = 1}^{N_P} T_{i,P} \cdot \frac{1}{K_{X,P}}\,, \label{average-time}
\end{equation}
where the sum runs over all the total duration times $T_{i,P}$ of the $N_P$ simulations and $K_{X,P}$ 
gives the total number of events $X$ in the set $P$. Every simulation has been given a burn-in time 
of $10^4 \yrs$, which is of the order of $\tau_{\rm RR} \approx 6 \times 10^4 \yrs$ (see Eq. \ref{eq.rr}) in order to discard merger events due to 
particles being created extremely close to the MBH or even within the capture radius by the 
randomization routine.

In Eq.~(\ref{average-time}), the possible events $X$ can be either $i$ for an inspiral event or $p$ 
for a plunge event.  We define the error for our results as the Poisson error,

\begin{equation}
 \sigma_{X,P} = \sqrt{K_{X,P}}\,.
\end{equation}
The criterion for an event to be an inspiral event is taken to be
\begin{enumerate}
\item ${\rm a}_{\rm cap} < 1\,{\rm mpc}$ and
\item ${\rm a}_{\rm cap} < 1.5\, {\rm a}(t_{\rm cap}-500\,{\rm yr})$\,.
\end{enumerate}
The second condition ensures that the semi-major axis of the merging body (${\rm a}_{\rm cap}$) has shrunk 
significantly prior to capture in order to dismiss plunges with low semi-major axis. The choice of $500$\,years
has empirically proven to distinguish perfectly between plunges and inspirals.

\subsection{Newtonian Simulations (Set SI)}

In this section we present the results for our study using only purely Newtonian accelerations, 
i.e. only the term $F_{0}$ in Eq.~(\ref{eq.F_PN}). 
In this case, the individual objects exchange energy and angular momentum efficiently via resonant 
relaxation (RR).  For the plunge time, using Eq.~(\ref{average-time}), we find

\begin{align}
\tau_{\rm p, I} = (3.7 \pm 0.2) \times 10^4\yrs\,.
\end{align}
This agrees very well with the RR timescale given by Eq.~(\ref{eq.rr}) for a typical particle, which 
confirms that this is the dominant mechanism at these radii for driving stellar objects into the central body in the 
purely Newtonian case. Of course, we do not identify any inspirals in the absence of gravitational 
radiation effects.

\subsection{Simulations including $2.5$PN corrections (Set SII)}
We now add only the dissipative effects due to gravitational radiation emission, which appear at 2.5 PN 
order, to the acceleration equations.

The analysis of the simulations now gives

\begin{align}
\tau_{\rm p, II} &= (3.8 \pm 0.2) \times 10^4\yrs \,,\\
\tau_{\rm i, II} &= (2.3 \pm 0.4) \times 10^5\yrs \,.
\end{align}
Now, with the inclusion of the effects of gravitational radiation, gradual inspirals into the MBH 
are possible.  This converts a subset of the plunge events from the simulations in set SI into inspiral 
events.  However, the inspiral time $\tau_{\rm GW}$ given by Eq.~(\ref{eq.peters}) only becomes smaller than 
the RR time for 
\begin{equation}
a(1-e) \lesssim 5\,r_g\,,  \label{eq.rrdecouple}
\end{equation}
which is smaller than the assumed capture radius. Thus, the efficient RR still drives the majority of 
particles into the capture radius before they can decouple from the stellar background and 
undergo a clean inspiral. In other words, the transfer of angular momentum to more eccentric orbits 
by RR is faster than the circularization by the dissipative 2.5PN term.

\subsection{Simulations including $1$PN, $2$PN and $2.5$PN corrections (Set SIII)}
In this set of simulations we include all PN terms up to 2.5PN order in our calculations. This introduces 
prograde Schwarzschild precession in addition to the dissipation produced by gravitational-wave emission. 
This effect is expected to increase the associated times for inspiral 
and plunges, since it eliminates efficient RR at high eccentricities. We find
\begin{align}
\tau_{\rm p,III} &= (1.3 \pm 0.2) \times 10^6\yrs \,, \\
\tau_{\rm i,III} &= (2.0 \pm 0.4) \times 10^6\yrs \,.
\end{align}
As expected, the inspiral and plunge times are now of the order of a two-body relaxation time, 
Eq.~(\ref{eq.nr}). We also see that the substantial difference between plunge and inspiral times seen in 
SII vanishes, because now the relaxation time-scale is much higher and gravitational radiation can 
more easily decouple the star from the stellar background. Compared to the previous set now the 2.5PN 
term is able to circularize and shrink the BH orbit faster than classical relaxation increases eccentricity.

\subsection{Simulations including $1$PN and $2$PN corrections (Set SIV)}
In order to compare our PN results with the results using geodesic equations of motion, we also ran a set of 
simulations with only the conservative 1PN and 2PN terms. 
In this set we find a plunge time of
\begin{equation}
\tau_{\rm p,IV} = (6.6 \pm 0.5) \times 10^5\yrs\,.
\end{equation}
This is consistent with the time we would obtain from SIII when combining plunge and inspiral events 
and shows again that the $2.5$PN term does not change the important mechanisms for angular momentum transfer.

\subsection{Simulations considering geodesic motion around the MBH (Set SV)}
In this set of simulations we investigate the system using Newtonian forces to describe the gravitational
interactions between the stellar black holes
and to describe the interaction between the MBH and individual stellar black holes we use 
the exact solution of the motion of a test 
mass in a Schwarzschild metric (geodesic motion). This does intrinsically exclude dissipative effects and 
therefore, the results of this subsection should be compared to those from the set SIV. 
In the limit of $m_\star \ll M_\bullet$, the geodesic equations (in harmonic coordinates) 
expanded up to $2$PN order and the conservative 2PN equations agree (see Appendix \ref{sec.appendix}), 
and hence they are consistent descriptions at this level of approximation. 

In this set of simulations we obtain

\begin{equation}
\tau_{\rm p,V} = (7.1 \pm 0.7) \times 10^5\yrs,
\end{equation}
which is consistent with $\tau_{\rm p,IV}$. This agreement means that the motion very close to the 
MBH is not relevant for the relaxation processes that drive BHs into plunge orbits.

\section{Schwarzschild barrier}
\label{sec.sb}

\begin{figure}
\resizebox{\hsize}{!}
          {\includegraphics[scale=1,clip]{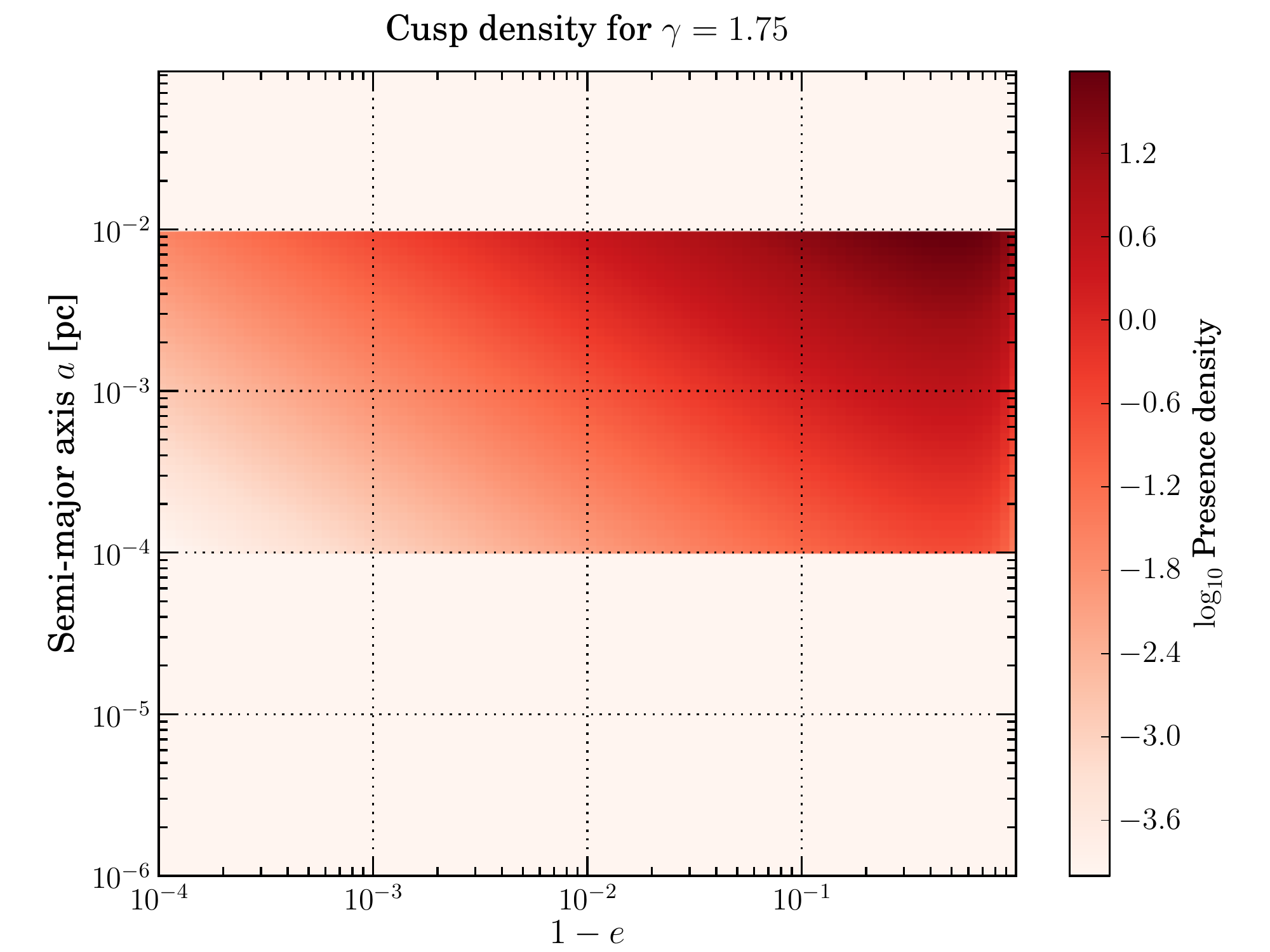}}
\caption
   {  
Theoretical distribution of a truncated cusp of $\gamma=1.75$
   }
\label{fig.cusp2d-v2}
\end{figure}

\begin{figure}
\subfloat[]{\includegraphics[width=\columnwidth]{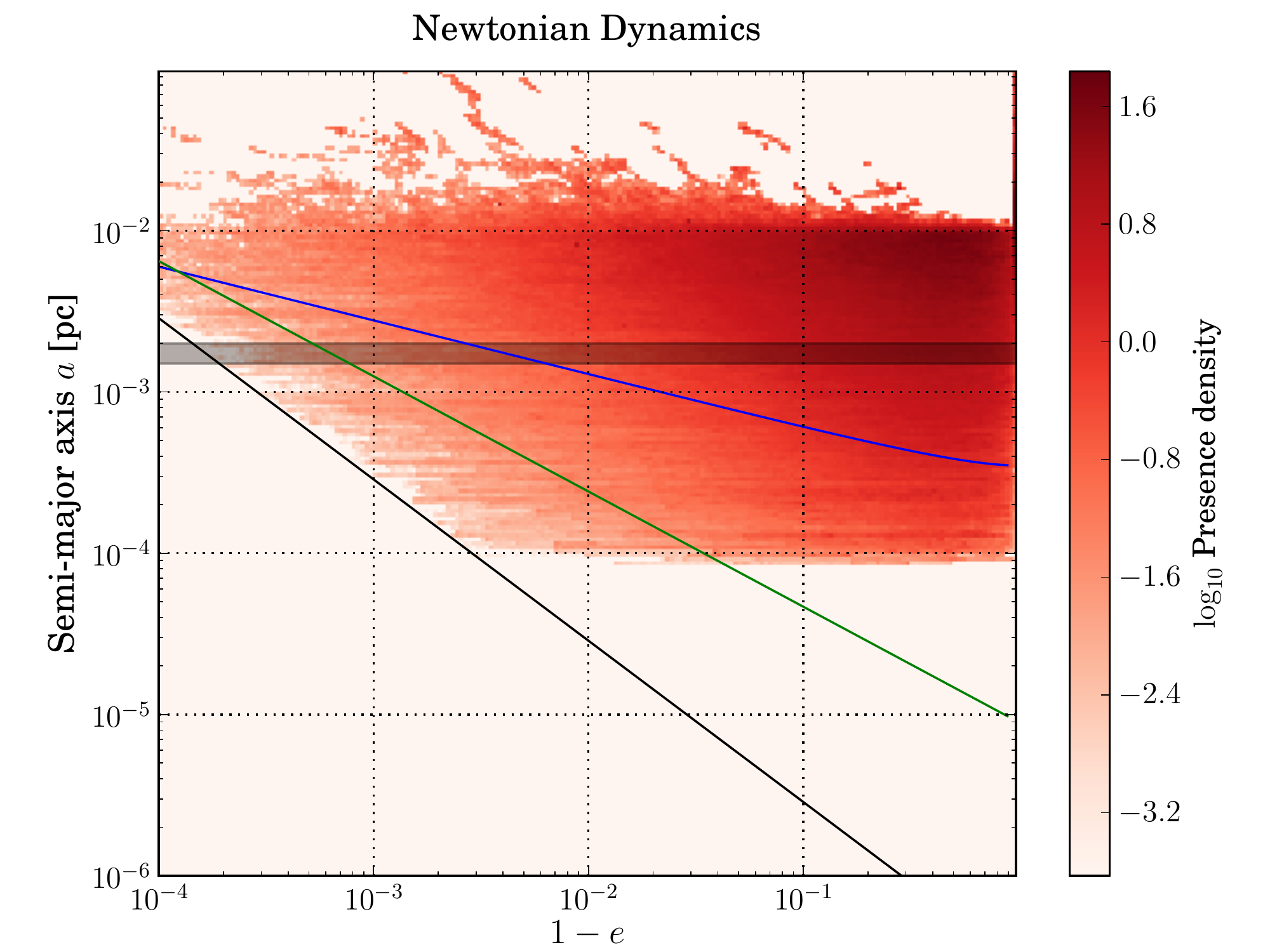}}\\
\subfloat[]{\includegraphics[width=\columnwidth]{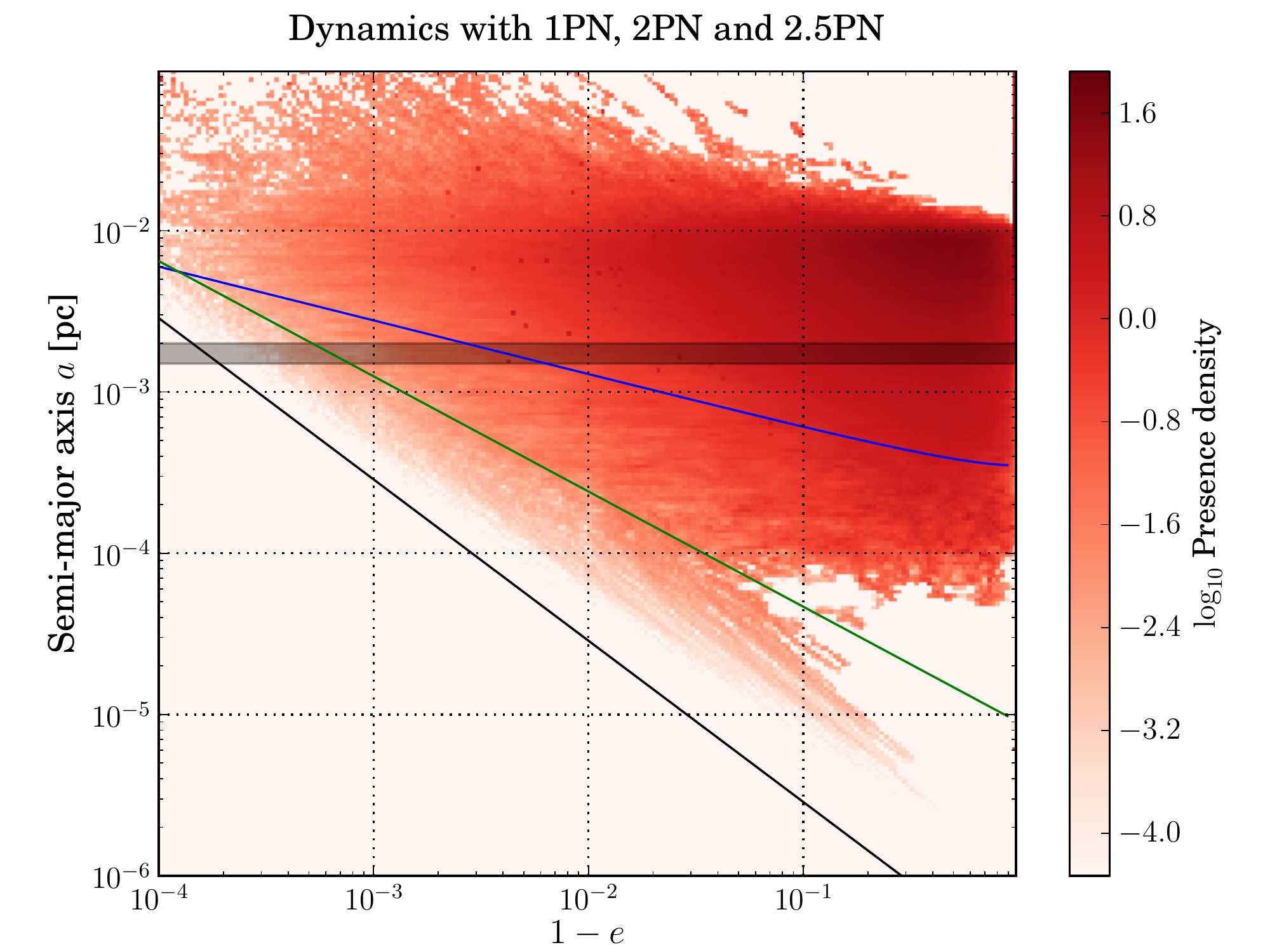}}
\caption
{Integrated presence density for the Newtonian (set SI, top panel) and the relativistic case 
(set SIII, bottom panel). The shaded box marks the region of the slice analyzed in Fig. \ref{fig.dndx}. 
The lines indicate the position of the Schwarzschild barrier with $C_{\rm SB} = 0.35$ ({\itshape blue}) 
and the limit for capture onto inspiral orbits for non-resonant relaxation ({\itshape green}).}
\label{fig.histo}
\end{figure}

\begin{figure}
{\includegraphics[width=\columnwidth]{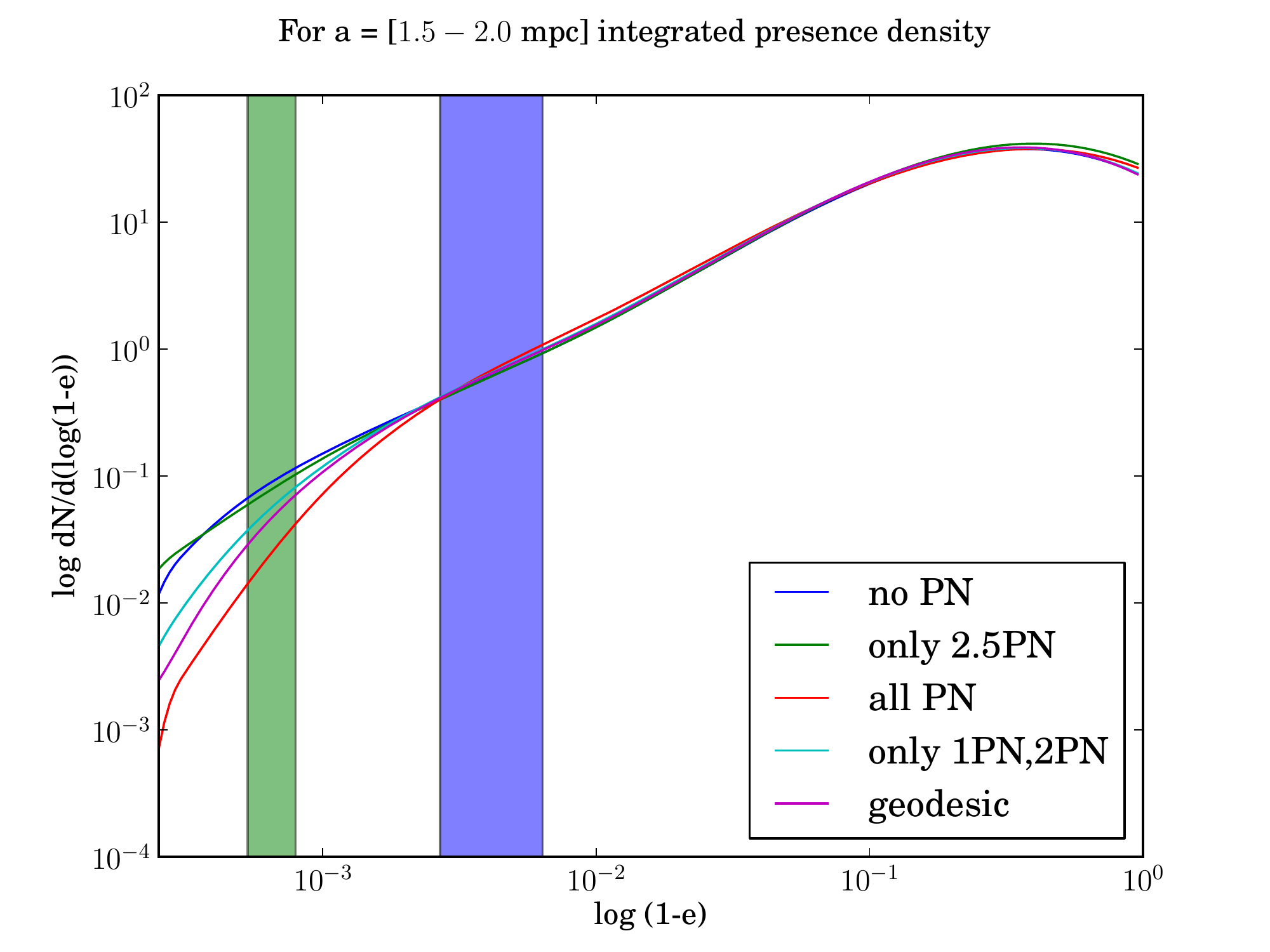}}
\caption
{Integrated presence density for a semi-major axis slice $1.5\,{\rm mpc} < a < 2.0\,{\rm mpc}$. The 
shaded boxes represent the position of the Schwarzschild barrier with $C_{\rm SB} = 0.35$ 
({\itshape blue}) and the threshold between the regime in which the evolution is dominated
by dynamics and when the binary decays via gravitational loss 
({\itshape green})}.
\label{fig.dndx}
\end{figure}

In order to quantify the nature of the ``Schwarzschild barrier'', we first plot
the normalized presence density. This is a measure of the total time any
particle spends in a certain logarithmic bin in $(a,1-e)$, summed up over all
particles and simulations. For comparison with the next figures, in figure \ref{fig.cusp2d-v2} we depict the
theoretical distribution that we can expect from a power-law of exponent $\gamma=1.75$. Next, we show the histogram for the Newtonian case, Fig.
\ref{fig.histo} (top panel) and the relativistic case (bottom panel). If we
consider our specific setup, there are 3 different regions in the $(a,1-e)$
plane where different mechanisms are efficient. In the right-most region, where
pericentres are large, RR plays the dominant role. The left border of this
region is roughly given by the appearance of the Schwarzschild precession which
inhibits the BHs from experiencing coherent torques. Following the derivation
in MAMW11, the time-scale for changes in angular momentum due to an enclosed
distribution of stars with mass $m_\star$ acting as a coherent torque is

\begin{equation} 
\tau_{\rm coh} = \frac{M_\bullet}{\sqrt{N(a)} m_\star}
\Bigl(\frac{a^3(1-e^2)}{G M_\bullet}\Bigr)^{1/2}\,.  
\end{equation}
The number of stars within a sphere delimited by the BHs semi-major axis,
$N(a)$ is related to the density profile $\rho (r)$. For a general power law
$\rho (r) \propto  r^{- \gamma}$, the number of stars within a certain radius
$a$ becomes
\begin{equation} 
N(r<a) = N_{<1} \tilde{a}^{3-\gamma}\,, 
\end{equation}
where $N_{<1}$ is the number of stars within a sphere of radius $1$ mpc.  The
condition for the Schwarzschild barrier is that the relativistic precession
time-scale, Eq.~(\ref{eq.ss}), becomes smaller than $\tau_{\rm coh}$, i.e.
\begin{equation} 
a (1-e^2)^{1/2} = \frac{3 G}{2 \pi c^2} 
\frac{M_\bullet^2}{m_\star} \sqrt{N(r<a)}\,.  
\end{equation}
In our model with $\gamma = 2$ and $N_{<1} \approx 5$, we obtain the relation
\begin{equation} 
N(r<a) \approx 5\,\tilde{a}\,, 
\end{equation}
and thus the barrier at 
\begin{equation} 
\tilde{a}_{\rm SB} \approx C_{\rm SB} (1-e^2)^{-1/3}\,,
\end{equation}
where $C_{\rm SB} \approx 0.35$ in this particular order of magnitude
comparison. This line is shown in blue in Fig.~\ref{fig.histo}.  The
bottom-left side of it is the region where RR is inefficient. However, NR is
still in play. The next delimiter is placed by the inspiral time-scale
$\tau_{\rm GW}$. As soon as $\tau_{\rm GW} < \tau_{\rm NR}$, BHs decouple from
the stellar background and inspiral gradually, driven by energy loss through
gravitational radiation. The condition for this, using Eq.~(\ref{eq.peters}) and
Eq.~(\ref{eq.nr}), yields
\begin{align} 
\nonumber \tilde{a}_{\rm GW} \approx &1.5 \times 10^{-2}
\Bigl(\frac{M_\bullet}{10^6 \msun}\Bigr) \Bigl(\frac{m_\star}{50
\msun}\Bigr)^{-2/7}\\ &\times \Bigl(\frac{N_{<1}}{5}\Bigr)^{-2/7}
(1-e^2)^{-5/7}\,.  
\end{align}
This line for gravitational capture against non-resonant relaxation is shown in
green in Fig.~\ref{fig.histo}.  The relativistic set (bottom panel) shows the
characteristic inspiral lines, which cause the presence very close to the
merger limit ($a < 6\,r_g$) to be depleted in comparison to the Newtonian case,
where the particles are scattered into plunge orbits onto the MBH instead.

Fig.~\ref{fig.dndx} shows the presence integrated over a small slice in semi-major axis (the shaded area in 
Fig.~\ref{fig.histo}) for all the simulated sets. In this plot we show the location of the barrier at 
$C_{\rm SB} = 0.35$, shown as the blue region. Since we average over a certain part in the semi-major axis,
the lines from Fig.~\ref{fig.histo} now appear as regions marking the lower and upper boundary at the limiting
semi-major axis values.

The same decrease in presence density can be observed for the runs with no dissipative forces, set SIV and SV, 
which both show the same behavior towards high eccentricities. This confirms the validity of the PN 
approximation in this regime for our purposes.  It also suggests that the system barely
access the regime where the PN and geodesic dynamics differ, i.e. where the relativistic
precession predicted by the geodesic equations of motion is significantly different from
the 1PN and 2PN corrections.  

The drop for set SIII as compared to IV and V is stronger, because of the presence of inspirals depleting 
the very high eccentricity region.

Set SI and SII also show consistent results. This was expected because the decoupling into inspiral orbits 
happens on length scales below the merger criterion for resonant relaxation, see Eq.~(\ref{eq.rrdecouple}).

\begin{table}
  \begin{tabular}{|c|c|c|}
    \hline
    Series & Plunge (yr$^{-1}$) & Inspiral (yr$^{-1}$)\\ \hline \hline
    SI: Newtonian & $(2.7 \pm 0.2) \times 10^{-5}$ & -- \\ \hline
    SII: Only 2.5 PN & $(2.6 \pm 0.2) \times 10^{-5}$& $(4.3 \pm 0.6) \times 10^{-6}$\\ \hline \hline
    SIII: Full PN & $(8 \pm 1) \times 10^{-7}$& $(5 \pm 1) \times 10^{-7}$\\ \hline \hline
    SIV: 1PN, 2PN & $(1.5 \pm 0.1) \times 10^{-6}$ & -- \\ \hline
    SV: Geodesics & $(1.4 \pm 0.1) \times 10^{-6}$ & --\\ \hline

  \end{tabular}
  \caption{Comparison of the different event rates for the different scenarios studied.}
  \label{tab.eventrates}
\end{table}

We summarize the event rates for our idealized system in Table
\ref{tab.eventrates}.
Our rates are only meant to compare different sets of simulations and can only be read as absolute
values for a real galactic nucleus.

\section{Discussion}

Recently, \cite{MerrittEtAl11} estimated with a few direct-summation
$N-$body simulations expanded with a statistical Monte-Carlo study that the
``traditional EMRI'' event is markedly decreased by the presence of a blockade
in the rate at which orbital angular momenta change takes place. This so-called
``Schwarzschild barrier'' is a result of the impact of relativistic precession
on to the stellar potential torques.  Although the authors find that some
particles can penetrate the barrier, EMRIs are significantly suppressed in this
scenario. 

In this study we investigated the effects of relativistic corrections on the
event rates for EMRIs compared to Newtonian dynamics using a PN approach as
presented in the original work of \cite{KupiEtAl06} but also and for the first
time with the implementation of geodesic equations.   For this purpose, we ran
different sets of 500 simulations each, for combined durations of
$\sim 100 \, {\rm Myr}$ in order to obtain statistically solid results. We
confirm the quenching of RR in the presence of Schwarzschild precession, i.e.
the Schwarzschild barrier. Comparing full PN simulations (up to order $2.5$)
with the Newtonian ones, we find a ratio of the time-scales for the capture
(combined plunge and inspiral) of 

\begin{equation}
  \tau_{\rm GR} / \tau_{\rm Newtonian} \approx 21 \pm 7
\end{equation}

\noindent
and for the absolute value for our setup an EMRI event rate of $\lesssim 1\,
{\rm Myr}^{-1}$. This value serves as an order of magnitude estimate for
galactic centers similar to our idealized setup.

In order to investigate the validity of the barrier at high eccentricities and
very small pericentres, we have implemented the geodesic equations of motion
around the MBH as an alternative to the PN corrections.  We find that the
results we obtain from both methods are consistent although the relativistic
precession they predict is significantly different near the last stable orbit.
This suggests that the stellar dynamics of the systems we have studied does not
access significantly the regime where the dynamics as described by the geodesic
equations and the PN corrections is different.  Therefore we deem it necessary
to address the scenario to check whether these results are recovered when we
increase the number of stars $N$.  We hence plan on expanding the studied
system for usage with $N-$body codes. This will allow us to investigate
realistic galactic nuclei and thus make more precise statements about the
absolute event rates to be expected.

We corroborate the existence of the ``Schwarzschild barrier'' and its impact on
the rate at which orbital angular momenta change takes place {\em at these
distances}.  It is important to note that recently
\cite{Amaro-SeoaneSopuertaFreitag2012} proved that for Kerr MBHs, direct
plunges are in reality high-eccentricity EMRIs, and that the rate in enhanced
depending on the spin and inclination. Although our simulations probe only an
idealized case in which we study a peculiar, though representative
distribution, the consequences are clear: The event rate of EMRIs for a
LISA-like mission such as eLISA
\citep{Amaro-SeoaneEtAl2012,Amaro-SeoaneEtAl2012b} will be dominated by
high-eccentric EMRIs, which are not ``direct plunges'' if the central MBH is
spinning. On the other hand EMRIs produced in the region in which it was
believed that resonant relaxation could represent a significant enhancement in
the rates, will suffer a drastic cut in the rates and, hence, they will be only
a negligible contribution.

\section*{Acknowledgments}
This work has been supported by the Transregio 7
``Gravitational Wave Astronomy'' financed by the Deutsche
Forschungsgemeinschaft DFG (German Research Foundation).
The simulations were run on the ATLAS cluster of the Albert-Einstein-Institute Hannover.
CFS acknowledges support from the Ram\'on y Cajal Programme of the Spanish Ministry of Education 
and Science, contract 2009-SGR-935 of AGAUR,
and contracts FIS2008-06078-C03-03, AYA-2010-15709, and FIS2011-30145-C03-03 of MICCIN. 
We acknowledge the computational resources provided by the BSC-CNS
(contracts AECT-2012-1-0008 and AECT-2012-2-0014) and CESGA (contracts CESGA-ICTS-221 and ICTS-CESGA-234).

\begin{appendix}

\section{Geodesic equations for a particle orbiting a Black Hole}
\label{sec.appendix}

In this appendix we write down the geodesic equations of motion in a form that is suitable to be
included in a N-body code that uses a Newtonian-type formulation of the equations
of motion.  In order to compare results with the cases where PN corrections are used we
write the geodesic equations using harmonic coordinates for Schwarzschild, which are 
compatible with the harmonic gauge condition of PN theory. 

Since our particles represent stellar objects we need to consider the geodesics for 
massive particles (i.e. timelike geodesics).   Given our system of spacetime coordinates
$\{x^\mu\} = \{t,x^i\}$ ($\mu\,,\nu\,,\dots = 0-3$;\; $i\,, j\,, \ldots = 1-3$), 
a geodesic will be given by $\{x^\mu(\tau)\}$, where $\tau$ denotes the particle's proper time.
The components of the velocity vector are defined as
\begin{equation}
u^\mu = \frac{dx^\mu(\tau)}{d\tau} \,.
\end{equation}
This four-velocity vector satisfies:
\begin{equation}
g^{}_{\mu\nu}u^{\mu}u^{\nu} = - c^{2} \,, \label{unorm}
\end{equation}
where $g^{}_{\mu\nu}$ is the Schwarzschild metric in our coordinate system and $c$ denotes the speed of light.
Since we are interested in geodesics, the velocity vector must satisfy the 
following equation of motion~\citep{MisnerThorneWheeler1973}.
\begin{equation}
u^\nu\nabla^{}_\nu u^\mu = 0\,, \label{geqs}
\end{equation}
where $\nabla^{}_\mu$ denotes the canonical covariant derivative associated
with the spacetime metric $g_{\mu\nu}$.  Expanding this equation we have
\begin{equation}
\frac{du^\rho}{d\tau} + \Gamma^{\rho}_{\mu\nu}u^\mu u^\nu = 0\,, \label{egeqs}
\end{equation}
being $\Gamma^\rho_{\mu\nu}$ the Christoffel symbols associated with the
spacetime metric $g^{}_{\mu\nu}$.  They are given in terms of the metric by:
\begin{equation}
\Gamma^{\mu}_{\alpha\beta} = \frac{1}{2} g^{\mu\nu}\left(
\frac{\partial g^{}_{\alpha\nu}}{\partial x^{\beta}} + 
\frac{\partial g^{}_{\beta\nu}}{\partial x^{\alpha}} -
\frac{\partial g^{}_{\alpha\beta}}{\partial x^{\rho}} \right)\,.
\label{christoffel-symbols}
\end{equation}
Using the splitting of time and space we can write the velocity vector as follows:
\begin{equation}
\vec{\bm u} = u^t \frac{\partial}{\partial t} + u^i\frac{\partial}{\partial x^i}\,,
\end{equation}
where $\{u^t,u^i\}$ are the velocity components in the $\{t,x^i\}$ coordinate
system:
\begin{equation}
u^t = \frac{\partial t(\tau)}{\partial \tau}\,,~~~~~
u^i = \frac{\partial x^i(\tau)}{\partial \tau} \,.
\end{equation}
Therefore, on the trajectory of the particle we can write
\begin{equation}
u^i = \frac{dx^i(t)}{dt}\frac{\partial t}{\partial \tau} = v^i u^t \equiv \Gamma v^{i}\,,
\end{equation}
where $v^i$ are the spatial components of the velocity
\begin{equation}
v^i = \frac{dx^i(t)}{dt}\,,
\end{equation}
and $\Gamma$ is the general relativistic version of the special relativistic gamma
factor, which is given in terms of the components of the spatial
velocity and the metric tensor as:
\begin{equation}
\Gamma^{2} = -\frac{c^{2}}{g^{}_{tt} + 2g^{}_{ti}v^i + g^{}_{ij}v^iv^j}\,.
\label{gamma-factor}
\end{equation}
which, in the weak-field limit ($g^{}_{tt}\approx -c^{2}\,$, $g^{}_{ti}\approx
0\,$, $g^{}_{ij}\approx \delta^{}_{ij}\,$), has the usual
expression:
\begin{equation}
\Gamma^{2} \approx \frac{1}{1-\frac{v^{2}}{c^{2}}}\,, \qquad (v^{2}\equiv \delta^{}_{ij}v^{i}v^{j})\,.
\end{equation}

At this point, we can now adopt a Newtonian point of view by looking at 
the geodesic equations for the six quantities: $\{x^i(t),v^i(t)\}$,
that is, for the spatial coordinates and spatial velocity components.
They can be written as:
\begin{align}
\frac{dx^i}{dt} &= v^i\,, \label{geq1}  \\
\frac{dv^i}{dt} &= \,f^i_{g} \,, \label{geq2}
\end{align}
where the {\em forces}, $f^i_{g}$, are actually forces per unit mass, i.e. 
accelerations, since they should not depend on the mass of the body (according
to the equivalence principle).  Moreover, these specific forces depend
on the spacetime metric (and its first derivatives) and on $v^i$.
They can be written as
\begin{align}
\nonumber f^i_{g} =&
v^i\,\Gamma^t_{tt}-\Gamma^i_{tt}
+ 2\left(v^i\,\Gamma^t_{tj}-\Gamma^i_{tj}\right)v^j\\
&+ \left(v^i\,\Gamma^t_{jk}-\Gamma^i_{jk}\right)v^jv^k\,. \label{force}
\end{align}
Given initial conditions $\{x^i_o,v^i_o\}$ equations (\ref{geq1},\ref{geq2})
have a unique solution $\{x^i(t),v^i(t)\}\,$.   
Note that the $c^{2}$ factor dividing
the forces, when going to the right-hand side of the equation (multiplying the 
Christoffel symbols) will cancel the $c^{2}$ factor in the denominator of $r^{}_{g}\,$
[see expressions in Eqs.~(\ref{gamma-t-tt})-(\ref{gamma-i-jk})].

Since up to now the development has been quite general, let us now consider the case of 
a non-spinning (Schwarzschild) MBH black hole of mass $M_{\bullet}$.  The metric components, 
in harmonic coordinates, can be written in the following form:
\begin{align}
g^{}_{tt} & =  - \frac{1-\frac{r^{}_{g}}{r}}{1+\frac{r^{}_{g}}{r}}\, c^{2} \,, \\[1mm]
g^{}_{ti} & = 0\,, \\[1mm]
g^{}_{ij} & =  \frac{1+\frac{r^{}_{g}}{r}}{1-\frac{r^{}_{g}}{r}}\, n^{}_{i}n^{}_{j}
+\left(1+\frac{r^{}_{g}}{r}\right)^{2}\left(\delta^{}_{ij} - n^{}_{i}n^{}_{j}\right) \,,
\end{align}
where
\begin{equation}
r = \sqrt{\delta^{}_{ij}\,x^{i}x^{j}}\,, \qquad
n^{i} = \frac{x^{i}}{r}\,,\qquad
r^{}_{g} = \frac{GM_{\bullet}}{c^{2}}\,.
\end{equation}
>From here, the components of the inverse metric are:
\begin{align}
g^{tt} & =  - \frac{1+\frac{r^{}_{g}}{r}}{1-\frac{r^{}_{g}}{r}}\,\frac{1}{c^{2}} \,, \\[1mm]
g^{ti} & = 0\,, \\[1mm]
g^{ij} & =  \frac{1-\frac{r^{}_{g}}{r}}{1+\frac{r^{}_{g}}{r}}\, n^{i}n^{j}
+\frac{1}{\left(1+\frac{r^{}_{g}}{r}\right)^{2}}\left(\delta^{ij} - n^{i}n^{j}\right) \,,
\end{align}
where $x^{}_{i}=\delta^{}_{ij}\,x^{j}$ and $n^{}_{i} =\delta^{}_{ij}\,n^{j}$.

The important thing to determine the forces is the computation of the Christoffel symbols.
>From their definition~(\ref{christoffel-symbols}) we find the following result
\begin{align}
\Gamma^{t}_{tt} = & 0 \,, \label{gamma-t-tt} \\[2mm]
\Gamma^{t}_{ti} = & \frac{r^{}_{g}}{r^{2}}\frac{n^{}_{i}}{1-\left(\frac{r^{}_{g}}{r}\right)^{2}} \,, 
\label{gamma-t-ti} \\[2mm]
\Gamma^{t}_{ij} = & 0 \,, \label{gamma-t-ij} \\[2mm]
\Gamma^{i}_{tt} = & \frac{r^{}_{g}}{r^{2}}\frac{1-\frac{r^{}_{g}}{r}}{\left(1+\frac{r^{}_{g}}{r}\right)^{3}}\,n^{i}\,c^{2} 
\,, \label{gamma-i-tt} \\[2mm]
\Gamma^{i}_{tj} = & 0\,, \label{gamma-i-tj} \\[2mm]
\nonumber \Gamma^{i}_{jk} = & \frac{r^{}_{g}}{r^{2}}\frac{1}{1+\frac{r^{}_{g}}{r}} \Bigl[ \left(1+\frac{r^{}_{g}}{r}\right)\,n^{i}
\left(\delta^{}_{jk}-n^{}_{j}n^{}_{k}\right) \\[2mm]
&- \frac{n^{i}n^{}_{j}n^{}_{k}}{1-\frac{r^{}_{g}}{r}}
-2 n^{}_{(j}\left(\delta^{i}_{k)}- n^{i}n^{}_{k)} \right) \Bigr] \,. \label{gamma-i-jk}
\end{align}
And this determines completely the geodesic equations of motion in Eqs.~(\ref{geq1}) and~(\ref{geq2}).

Finally, we can make a post-Newtonian expansion of the equations of motion.  That is, an expansion
for $r^{}_{g}/r \ll 1\,$, and $v/c \ll 1\,$.  In our case, the expression for the {\em force} 
simplifies to [see Eq.~(\ref{force}) and Eqs.~(\ref{gamma-t-tt})-(\ref{gamma-i-jk})]: 
\begin{equation}
f^i_{g} = -\Gamma^i_{tt} + 2\,v^i\,\Gamma^t_{tj}v^j
-\Gamma^i_{jk}v^jv^k\,. \label{force2}
\end{equation}
Expanding this we get:
\begin{align}
f^{i}_{g} = & -\frac{r^{}_{g}c^{2}}{r^{2}}\left[ 1 - 4\frac{r^{}_{g}}{r} + 9\left(\frac{r^{}_{g}}{r}\right)^{2} 
- 16\left(\frac{r^{}_{g}}{r}\right)^{3} \right]\,n^{i} \nonumber \\[2mm]
&+ 2\,\frac{r^{}_{g}c^{2}}{r^{2}} \left[ 1+\left(\frac{r^{}_{g}}{r}\right)^{2}\right] \left(\frac{n^{}_{j}v^{j}}{c}\right)
\frac{v^{i}}{c} \nonumber \\[2mm]
& -\frac{r^{}_{g}c^{2}}{r^{2}} \Bigl\{ \,n^{i}\left(\delta^{}_{jk}-n^{}_{j}n^{}_{k}\right) 
- \left[ 1 + \left(\frac{r^{}_{g}}{r}\right)^{2} \right] n^{i}n^{}_{j}n^{}_{k} \nonumber \\
&- 2 \left[ 1 - \frac{r^{}_{g}}{r} + \left(\frac{r^{}_{g}}{r}\right)^{2} - \left(\frac{r^{}_{g}}{r}\right)^{3} \right] 
n^{}_{(j}\left(\delta^{i}_{k)}- n^{i}n^{}_{k)} \right) \Bigr\} \nonumber \\
& \times \frac{v^{j}}{c}\frac{v^{k}}{c}\,,
\end{align}
where the first two rows correspond to the first two terms in Eq.~(\ref{force2}).  We have expanded in Taylor
series the functions of $r^{}_{g}/r$ up to order $(r^{}_{g}/r)^{4}\,$.  We can now collect the terms and
we find the following expression, which is valid to order 2PN [see Eq.~(\ref{PN_force}) below]:
\begin{align}
f^{i}_{g} = &- \frac{GM_{\bullet}}{r^{2}}n^{i} 
+ \frac{GM_{\bullet}}{r^{2}}\Bigl\{ \left({\cal A}^{}_{\rm 1PN} + {\cal A}^{}_{\rm 2PN}\right)n^{i} \nonumber \\
&+ \frac{\bm{n\cdot v}}{c} \left({\cal B}^{}_{\rm 1PN} + {\cal B}^{}_{\rm 2PN}\right)\frac{v^{i}}{c}\Bigr\}\,,
\label{geodesic-PN_force}
\end{align}
where
\begin{align}
\frac{\bm{n\cdot v}}{c} = &\frac{\bm{x}}{cr}\frac{d\bm{x}}{dt} = \frac{1}{2cr}\frac{d\bm{x}^{2}}{dt} = \frac{1}{2cr}\frac{dr}{dt}
= \frac{\dot{r}}{c} \,, \nonumber \\
v^{2} = &\bm{v\cdot v} = \delta^{}_{ij}v^{i}v^{j}\,,
\end{align}
and
\begin{align}
\label{eq.carlos1}
{\cal A}^{}_{\rm 1PN}  = & 4\frac{r^{}_{g}}{r} - \frac{v^{2}}{c^{2}}\,, \\[1mm]
{\cal A}^{}_{\rm 2PN}  = & -9\left(\frac{r^{}_{g}}{r}\right)^{2} + 2 \left(\frac{\bm{n\cdot v}}{c}\right)^{2}\frac{r^{}_{g}}{r}\,, \\[1mm]
{\cal B}^{}_{\rm 1PN}  = & 4\,, \\[1mm]
{\cal B}^{}_{\rm 2PN}  = & -2\frac{r^{}_{g}}{r}\,. 
\label{eq.carlos2}
\end{align}

\section{PN corrections} \label{pn-corrections}
The PN equations of motion used in our simulations can be written in the 
form given in Eq.~(\ref{eq.F_PN}).  We can organize the different terms
in the following form (which is similar to the one used above in Eq.~(\ref{geodesic-PN_force}) for 
geodesic equations):
\begin{align}
f^{i}_{g} = &- \frac{GM}{r^{2}}n^{i} + \frac{GM}{r^{2}}\Bigl\{ \left({\cal A'}^{}_{\rm 1PN} + {\cal A'}^{}_{\rm 2PN}\right)n^{i}
 \nonumber \\[1mm]
+& \frac{\bm{n\cdot v}}{c} \left({\cal B'}^{}_{\rm 1PN} + {\cal B'}^{}_{\rm 2PN} \right)\frac{v^{i}}{c}\nonumber\\[1mm]
+& \frac{\bm{n\cdot v}}{c}{\cal A'}^{}_{\rm 2.5PN}\;n^{i} + {\cal B'}^{}_{\rm 2.5PN}\frac{v^{i}}{c}\Bigr\}\,,
\label{PN_force}
\end{align}
where here $M=m_{\star}+M_{\bullet}$ is the two-body (MBH+star) total mass. 
We list here the PN coefficients [see, e.g.~\citep{Blanchet06}, Eq.~(131)] for $m_\star \ne 0$:
\begin{align}
\label{eq.PNbegin}
{\cal A'}^{}_{\rm 1PN} = & \frac{3}{2}\nu\left(\frac{\bm{n\cdot v}}{c}\right)^{2} -(1+3\nu)\frac{v^{2}}{c^{2}}  + \left(4+2\nu\right)\frac{R^{}_{g}}{r}\,,\\[1mm]
{\cal A'}^{}_{\rm 2PN} = & -\frac{15}{8}\nu\left(1+3\nu\right)\left(\frac{\bm{n\cdot v}}{c}\right)^{4} \nonumber \\[1mm]
                       + &  \nu\left(3-4\nu\right)\left[\frac{3}{2}\left(\frac{\bm{n\cdot v}}{c}\right)^{2}-\frac{v^{2}}{c^{2}}\right]\frac{v^{2}}{c^{2}} \nonumber \\[1mm]
                       + & \frac{R^{}_{g}}{r} \Bigl\{ 2\left(1+\frac{25}{2}\nu+\nu^{2}\right)\left(\frac{\bm{n\cdot v}}{c}\right)^{2} \nonumber\\[1mm]
                         & +\nu\left(\frac{13}{2}-2\nu\right) \frac{v^{2}}{c^{2}}\Bigr\} 
                       - \left(9+\frac{87}{4}\nu\right)\frac{R^{2}_{g}}{r^2}\,, \\[1mm]
{\cal A'}^{}_{\rm 2.5PN} = &  \frac{24}{5}\frac{R^{}_{g}}{r}\frac{v^2}{c^{2}} + \frac{136}{15}\nu\left(\frac{R^{}_{g}}{r}\right)^{2}\,,\\[1mm]
{\cal B'}^{}_{\rm 1PN} = & 4-2\nu\,,\\[1mm]
{\cal B'}^{}_{\rm 2PN} = & -\frac{3}{2}\nu\left(3 +2\nu\right)\left(\frac{\bm{n\cdot v}}{c}\right)^{2} + \nu\left(\frac{15}{2}+2\nu\right) \frac{v^2}{c^{2}} \nonumber \\[1mm]
                         &-\left(2+\frac{41 \nu}{2} +4 \nu^2\right)\frac{R^{}_{g}}{r}\,,\\[1mm]
{\cal B'}^{}_{\rm 2.5PN} = & -\frac{24}{5}\nu\left(\frac{R^{}_{g}}{r}\right)^{2} -\frac{8}{5}\nu\frac{R^{}_{g}}{r}\frac{v^{2}}{c^{2}}\,.
\label{eq.PNend}
\end{align}
where $\nu$ is the symmetric mass ratio, $\nu = m_\star M_\bullet/M^2$, and
$R^{}_{g} = GM/c^{2}$. 
One can verify that the coefficients in Eq.~(\ref{eq.carlos1}) to Eq.~(\ref{eq.carlos2}) 
agree with Eq.~(\ref{eq.PNbegin}) to (\ref{eq.PNend}) for $\nu = 0$.

\end{appendix}

\label{lastpage}

\label{lastpage}
\end{document}